# Analyzing the Structure of Mondrian's 1920-1940 Compositions

Loe Feijs

Eindhoven University of Technology

`l.m.g.feijs@tue.nl`

**Abstract**. Mondrian was one of the most significant painters of the 20th century. He was a prominent member of DeStijl, the movement which revolutionized art by setting it free of the obligation to make images of existing objects, persons, or situations. DeStijl was one of the interrelated movements in early 20th century Europe including Cubism, Constructivism, the Futurists, and Dada. Their disruptive ideas changed the meaning of Western art. It was Mondrian, more than anyone else, who worked restlessly to find expression for the purest possible kind of beauty and truth, based on a theory called Neoplasticism. He was already famous during his lifetime and still now, his name is almost synonym for modern art. We analyze the structure of the system of black lines in his paintings and put the hypothesis to the test that the paintings could be obtained by recursive (binary) splitting. We used a novel tailor-made interactive analysis tool and apply it to as many Mondrian paintings as possible (in total 147). The results will be explained in a visual manner, but we also present statistical findings from the analysis of the 147 paintings. Our main conclusion is that the hypothesis of splitting decomposition is in general not true. It is possible to make Mondrian-like compositions by splitting, yet one misses out on a great deal of Neoplastic beauty if one would work by splitting only. It is possible to consider all crossings as pairs of Tees, but that is clumsy, and it leaves out essential information. Moreover there are other important design decisions of Mondrian, such as the keeping-distance to the canvas-edge which are not well-described by splitting.

**Q&A**. Can you give a precise definition of the splitting hypothesis? The splitting hypothesis is the idea that Mondrian's compositions can be obtained by recursive binary splitting of a white rectangle, marking each splitting by a solid black line and finally filling some of the un-split rectangles with gray or primary colors. Hereby it is understood that a crossing of a horizontal and a vertical line is not seen as two Tees at the same point (which would be a strange coincidence).

What statistical methods were used in the evaluation of the obtained data? That is, how was the hypothesis tested? The statistical methods used are applied after assigning a number to each painting, its splittingness in the range from zero to one, then calculating the percentage of paintings with splittingness 0 (2%) and splittingness 1 (10%), and the average splittingness (48.8%). We plot both splittingness and a simple complexity measure as a function of time and by visual inspection we note several trends. If we consider Piet Mondrian's 1920-1940 oeuvre as a sampling of an unknown ideal design space of worthwhile compositions, a one-sample Wilcoxon signed rank test shows that the median of the splittingness (in this design space) is not 100%, whereas a median of 50% cannot be rejected (significance level 0.05).

Can you formulate a final take home message? The take home message is that the splitting hypothesis is general not true and although it is possible to make Mondrian-like compositions by binary splitting, one misses out on a great deal of Neoplastic beauty if working by splitting only. Considering all crossings as pairs of Tees is clumsy and leaves out essential information while there are other important design decisions of Mondrian, such as the keeping-distance to the canvas-edge which are not well-described by splitting. There are options for future work such as quaternary splitting and entropy-based complexity measures.

## Who was Mondrian and why are his compositions important?

Piet Mondriaan was born in Amersfoort, The Netherlands in 1872. Later he changed his name into Mondrian. He worked in the Netherlands, in Paris, London, and New York. He was trained in traditional drawing and painting; he was talented and moderately successful at a national level.
In the years 1910-1920 his style changed in a revolutionary way, first being connected to the cubist movement, and later during the first world war as the international contacts were hampered, he worked with like-minded Dutch artists to develop the DeStijl movement (The Netherlands were neutral in the war). Then he moved to Paris again until in 1938, in view of the new war dangers, he fled to London and in 1940 to New York. There is a variety of biographies such as Seuphor (1960), Hanssen (2015) and Janssen (2016).

Mondrian was one of the most significant painters of the 20th century. He was a prominent member of DeStijl, the movement which revolutionized art by setting it free of the obligation to make images of existing objects, persons, or situations. DeStijl was one of the interrelated movements in early 20$^{th}$ century Europe including Cubism, Constructivism, the Futurists, and Dada. Their disruptive ideas changed the meaning of Western art. It was Mondrian, more than anyone else, who worked restlessly to find expression for the purest possible kind of beauty and truth. Together with Theo van Doesburg, he wrote about these ideas, developing a theory called Neoplasticism. He was already famous during his lifetime and still now, his name is almost synonym for modern art. Dozens of scholars are active researching his life and his work. The most prestigious museums of the world are proud to display his paintings.

Although most educated people have some understanding of what Mondrian's paintings look like, the impact of the work and the seriousness and devotion of the man demand that we as researchers try to experience and understand his paintings as thoroughly as possible. The most well-known paintings are compositions having a system of black lines, some of the rectangles in between being painted in primary colors. But what precisely is this system of black lines? Is there a simple or perhaps a more complex recipe behind it? Scholars have written about it in words (Blotkamp 1994, Schufreider 1997) whereas programmers have used computer languages to try and capture the essence of a possible system (Noll 1966, Feijs 2004, De Silva Garza & Lores 2004, Van Hemert and Eiben 1999). Most authors of Mondrian-generating computer programs use anecdotic evidence to support claims of results being Mondrian-like (comparing informally to a few selected compositions). Many programs can be found online for example Mondrimat by Stephen Linhart (www.stephen.com/mondrimat/), and Automondrian by Bruce Ediger (www.stratigery.com/automondrian.php).

In this paper we present novel analytical research looking at (almost) all compositions having a system of black lines, some of the rectangles in between being painted in primary colors. Nowadays we are in the fortunate circumstances that all known Mondrian paintings are catalogued (Joosten & Welch) and that since 2018 images of the paintings are available online (http://pietmondrian.rkdmonographs.nl/).

## The hypothesis of splitting decomposition

There is a rich variety of computer programs aimed at producing images with Mondrian-like qualities. The tradition of making such Mondrian generators began with Michael A. Noll in the 1960 and continues till today (Jacobs 2013, Nichol 2013, and Feijs 2019). There are various ways of producing the main structure of the painting, usually first making a kind of black-lines structure and then filling some of the resulting rectangles with primary colors. What is the best way to generate this structure is somewhat unclear, various authors and programmers explore different approaches.

One obvious idea is to use a recursive binary splitting procedure. Skrodzki and Polthier (2018) use recursive splitting, based on the KdTree data structure and they discuss examples of Mondrian's compositions where this works and where this does not work. Skrodzki and Polthier move on to make 3D artworks, which is great, but not testable against real Mondrian paintings. For some of Mondrian's compositions, the idea of recursive splitting really cannot work, in other cases it can, but then one has to treat two crossing black lines as if they are two Tee joints, which miraculously come together in the same place (here we call it a "strange coincidence". Even if it is a recursive splitting, the further splitting is not independent of the other split-off sibling sub-compositions. The idea of recursive splitting occurs in many places: it turns out to be a typical classroom exercise (see footnotes [1] [2] [3] [4]). Other approaches are more complex and more of a bottom-up nature (for example Feijs in Leonardo 2004 and Feijs in IJMA 2019) whereas it is also possible to use an evolutionary approach (for example Van Hemert and Eiben, 1999).

The main question in this article is to find out whether the recursive splitting procedure is a good way to understand the structure of Mondrian's compositions. If not, what are the obstacles that resist the proper working of recursive splitting? If some of the Mondrian's are recursive splittings and others are not, then the question is whether we can quantify the degree to which they are a recursive splitting.

## Analysis of selected compositions

First let have a look at B131: *Composition with red, yellow, black, blue, and gray* (1921). We have developed a special interactive analysis tool by which we insert green marking line segments to locate Mondrian's black lines. The tool works under the assumption that all lines are splitting lines: it will extend each marking line segment in both directions until it either hits an earlier line or the edge of the canvas. The marking line segments are green, because Mondrian did not use green, so the markings stand out. It is ugly but very practical for research purposes. The marking lines are not always precisely in the middle of the black line (the tool has a grid-snapping mechanism) but that is no problem, we look at the structure of the system of lines. This painting falsifies the hypothesis of splitting decomposition: there is no full-length horizontal or vertical line which splits the plane. In fact, most lines do not even touch the boundary of the canvas.

---

[1] www.integraldomain.org/lwilliams/mis370/classNotes/mondrian/mondrian.html
[2] people.reed.edu/~esroberts/archive/csci121-2019-Fall/handouts/30-TheRecursiveParadigm.pdf
[3] www.oreilly.com/library/view/thinking-recursively-with/9780471701460/9780471701460_mondrian_and_computer_art.html
[4] /nifty.stanford.edu/2018/stephenson-mondrian-art/

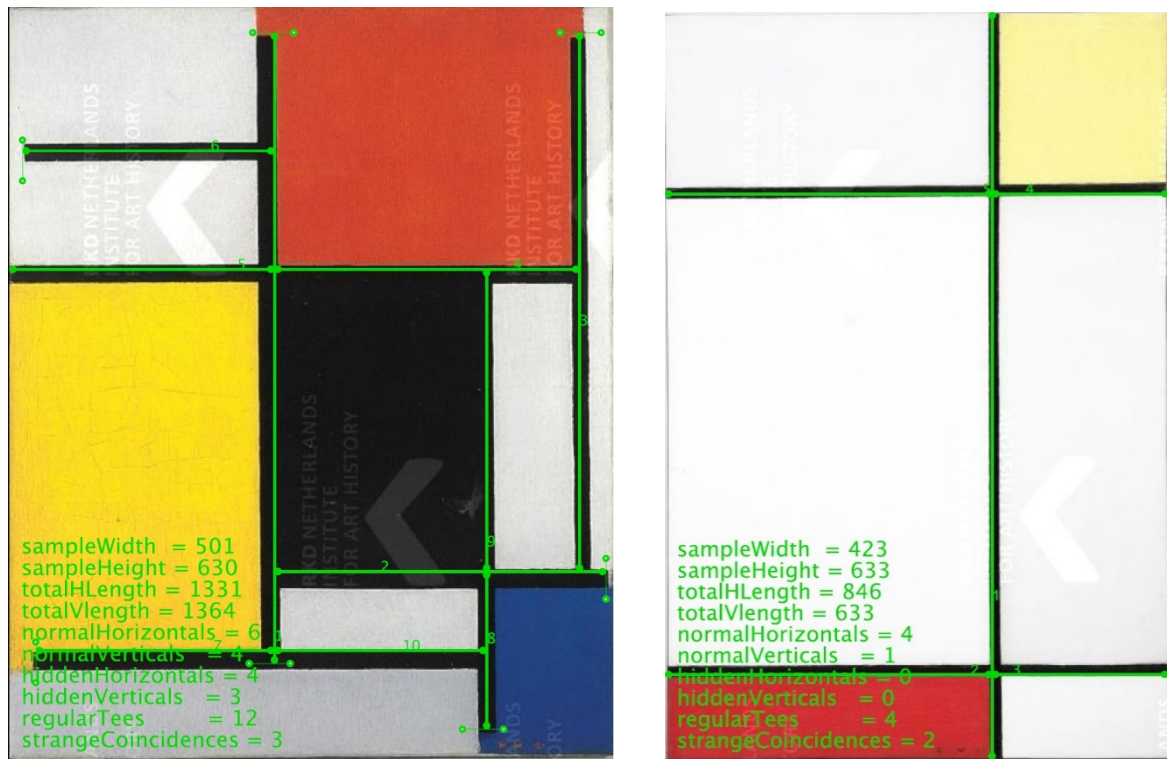

**Figure 1:** Paintings B131 and B198 with marking.

Let us deal with the last problem first: we add some so-called hidden lines to act as a stopping line for a longer line which would otherwise hit the boundary. They are also green, but thinner and of fixed length. If we ignore this problem, as if we would extend Mondrian's black lines near the edge of the canvas so as allowing them to run till the edge, then still the composition does not allow for a recursive splitting. In the tool with its built-in assumption of all lines being splitting lines, there is no way to make a correct first marking line segment. In order to perform our marking, we used a trick: one extra hidden marking line segment (the leftmost lowermost), which stops the long vertical.

We call the meeting point of two lines a *Tee*. Then we see a proper Tee where marking line 2 meets marking line 1. But we see a very particular situation where marking lines 4 and 5 meet 1. These form two Tees, but there is the strange coincidence that they meet *at the same point*. It would be common sense to say that line 1 and the combined line 4+5 cross each other. The hypothesis of splitting decomposition does not do justice to the real situation. We continue making markings as if all lines are splitting lines, but we count these *Strange Coincidences*, as we shall call them. The more Strange Coincidences, the weaker the status of the hypothesis of splitting decomposition. There are 12 regular Tees (i.e. we do not count the Tees involving hidden marking line segments) and 3 Strange Coincidences (which involve 6 regular Tees). So, even if we disregard the fact that one hidden line (the trick) is really far-off the edge of the canvas, 6/12 = 50% of the Tees are supportive of the splitting hypothesis; the others testify against it.

Next, consider B198, *Compositie with yellow and red* (1927). B198 can be marked without need for hidden marking lines, but it has all its Tees being part of a Strange Coincidence. It is beautiful, simple, just three crossing lines. Of course the beauty also arises from the balance, the precise positions of the lines, the colored planes, but that is beyond the formal structural analysis we conduct here. The author feels sorry for putting the ugly green lines over it.

For B125, *Compositie met rood, blauw, zwart, geel en* grijs (1921), we find that, besides the need for three hidden line segments near the edge, there are no Strange Coincidences. In that sense it supports the splitting hypothesis for 100% (besides the three hiddens).

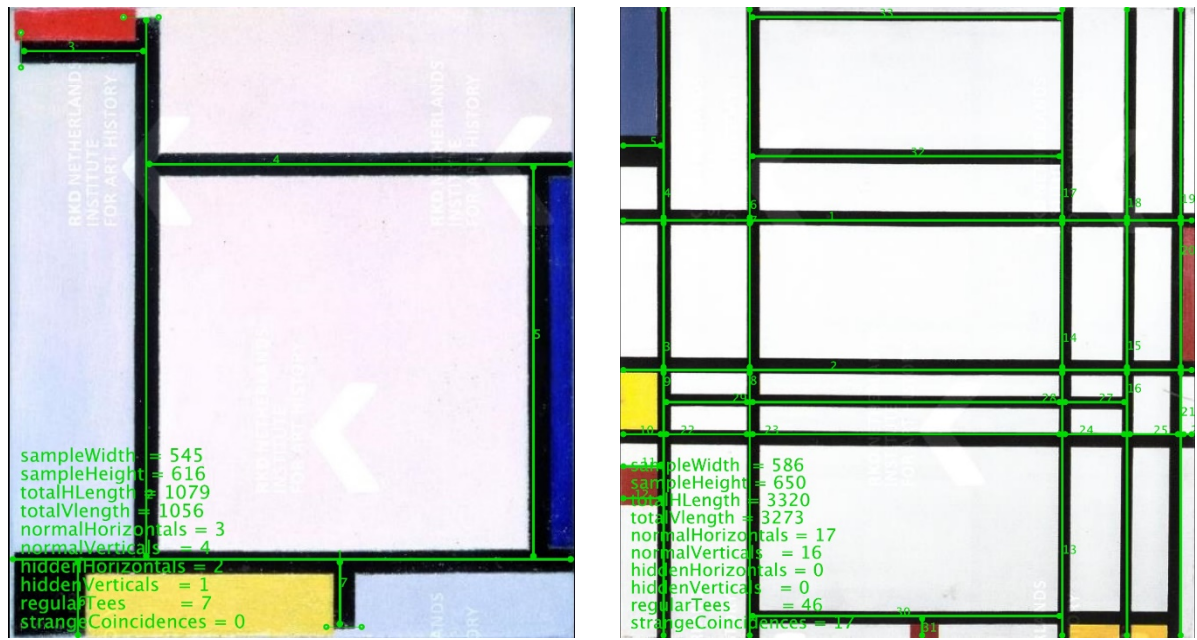

**Figure 2:** Paintings B125 and B288 with markings.

As a 4th selected composition, consider B288, *Composition No.10* (1938). It has 46 regular Tees, no hidden lines needed. It has 17 Strange Coincidences, so only 26% of the Tees are supportive of the splitting hypothesis. The author believes that it is precisely the interplay between some lines being proper splitters, others crossing each other, which makes this system of lines beautiful and interesting.

In B288 we see something else, namely Mondrian begins using something which is not exactly a line and also not exactly a colored plane. From 1940 onwards this feature becomes very dominant, and that delimits the applicability of the concept of a *system of black lines, some of the rectangles in between being painted in primary colors*. For B288 we made the somewhat arbitrary decision that the very thick black line is still a line, the leftmost red is a plane bounded by lines and the lower red is a line itself.

It is important to mention that the lines not running till the edge of the canvas is not some kind of anomaly or sloppiness (Mondrian was extremely precise). It is part of a deliberate strategy after 1920 to eliminate perceptual qualities in the painting such as depth or individual objects. This what Carel Blotkamp in his book calls *The art of Destruction* (Blotkamp 1994). There are other structural aspects in the painting which escape our line-based analysis, for example the phenomenon of superblocks (see B296), which are fully in line with Blotkamp's theory.

## Systematic analysis of all compositions
### Variables

The following variables are recorded for each painting. SW: sample Width (in pixels); SH: sample Height (in pixels); THL: total length of horizontals (in pixels); TVL: total length of verticals (in pixels); NH: number of normal (= non-hidden) horizontals; NV: normal verticals; HV: number of hidden horizontals; HH: idem verticals; RT: number of regular Tees; SC: number of strange coincidences. The numbers are reported by the tool after the interactive marking process.

From these we calculate SPLITTINGNESS, essentially the *percentage of Tees not part of a strange coincidence* as (RT ─ 2×SC) / RT. It is in the range from 0% to 100%. The higher this number, the more the painting supports the splitting hypothesis.

We also calculate a COMPLEXITY, a simple complexity measure, as (THL + TVL) / (SW + SH). A painting with one full length horizontal and one full length vertical line thus has simple complexity measure 1.0. The more lines, the higher the number.

Finally we calculate the SPECIAL EFFECTS, the number of boundary-distance-keeping tricks we had to deploy by using a hidden marked line segment, as HH + HV. If this 0, there is no problem, otherwise the splitting hypothesis is somewhat violated. Of course this is not a problem for the Mondrian painting, on the contrary, it is an important and well-justified design decision by Mondrian.

## Research questions

The top level research question is to find out to which degree the hypothesis of splitting decomposition holds. This research question is subdivided into three sets of more specific research questions (RQ):

(RQ1) in how many paintings is the hypothesis entirely true, in how many entirely false, and how is the distribution of *splittingness* for the mixed type?

Next, (RQ2) whether we see trends in *splittingness* and *complexity*?

Finally (RQ3), we want to see in how many paintings the hidden line segments are needed, and if so, how many hidden line segments. The more hidden line segments needed, the weaker the status of the hypothesis of splitting decomposition.

## Inclusion and exclusion criteria

We include all of Mondrian's paintings from 1920 (included) to 1940 (included). The year 1920 completes the transition from figuration to non-figuration. Perhaps the 1919 paintings B99-B103 could have been included, but they would not benefit much from our analysis, as B102-B103 are the checkerboards, and B99-B101 are obtained by sub-setting the lines of a checkerboard grid. Moreover, most pre-1920s do not yet have the red-yellow-blue color scheme yet. The limit of 1940 is chosen because around that time another transition happened, which blurred the distinction between lines and planes. Already for B288 we had difficult choices to make (during our analysis) when marking lines. In later works such as B300 (New York City 1, unfinished, 1941, we see crossing colored lines and the paradigm of "compositions having a system of black lines, some of the rectangles in between being painted in primary colors" is no longer applicable. Of course we faced the problem that Mondrian reworked his paintings over and over; this we coped with by simply following the Catalogue Raisonné from 1920 to 1940, which means the paintings B104 to B298.

From these 192 paintings we exclude the lozenges (B127, B151, B152, B156, B165, B169, B173, B176, B210, B218, B229, B241, and B282). The lozenges would be hard to analyze because many the Tees or crossings are imaginative, outside the canvas, or perhaps not existing at all. Furthermore we exclude the paintings of which only a skewed photograph exists (e.g. B280), incomplete sketches (B111, B158, B164, B290, B291, B297, B298), works which are no paintings at all (e.g. B205, with the "poème" text), and a few works which are almost copies of another work (e.g. B224, which is almost B223). Finally we omitted also a few 1938 works which, like many more after 1940, consist of color lines mostly. As a result of these decisions, 150 paintings are eligible for analysis.

## Data acquisition

All images were downloaded from the website http://pietmondrian.rkdmonographs.nl, which is based on the Catalogue Raisonné (Joosten & Welch 1996). The typical image size was around 500 or 600 pixels for the length and height. High resolution images were reduced. Frames etc. were cut off in a bitmap editor (so we would not have to deploy hidden marking line segments for improper reasons).

## Data processing

The images were loaded in MondrianAnalyser, a tailor-made tool written in Processing 3 (www.processing.org). This tool facilitates interactive adding of green marking line segments according to the hypothesis of splitting decomposition. The code comprises four files of 106 (mondrianAnalyser.pde), 62 (File.pde), 119 (Line.pde), and 122 (Stats.pde) lines of code respectively. Once a painting image is loaded in the background, the main interactions are: left mouse click (make a horizontal green marking line segment), right mouse click (idem vertical), key BACKSPACE (undo last action), key '-' (next line will be hidden), and key 's' (save statistics and marked .pdf file).

All interactive marking was done by the author. The main strategy used was to look for a full splitting line if possible, otherwise deploy a hidden line segment. Despite the clear strategy, there are still many arbitrary decisions. Fortunately, most of the arbitrary choices do not affect the variables, for example if there is a cross, one can do the vertical first and then two horizontals, or first the horizontal and then two verticals; but in both cases the count of RT (number of regular Tees) and SC (strange coincidences) is the same, viz. RT = 2 and SC = 1.

## Data analysis

The MondrianAnalyser tool saved the values for all ten variables in a file, one file for each painting e.g. B104.txt for B101.jpg. The ten values of each painting where copied into an Excel file (Microsoft Office Professional Plus 2013), which already contained the catalogue numbers, years, and the inclusion decision for all paintings B104 to B298. Excel formulas were written to calculate SPLITTINGNESS, COMPLEXITY, SPECIAL EFFECTS, and to count the various types of occurrences. We used IBM® SPSS® Statistics Version 25 to perform a One-Sample Wilcoxon Signed Rank Test for hypothesized median values of 1.0 and 0.5. Unlike a t-test, this test works for data whose distribution is not normal.

## Results

RQ1: SPLITTINGNESS has a mean of 48.8% (SD = 26.5%). In 3 (2%) paintings it is 0% (it is a full grid), in 15 (10%) paintings it is 100% (i.e. the painting is a recursively splittable decomposition). We find that the distribution of SPLITTINGNESS is not normal (one-sample Kolmogorov-Smirnov test, significance level 0.05). The one-sample Wilcoxon signed rank test shows that the median of the splittingness is not 1.0. It also shows that a median of 0.5 cannot be rejected (significance level 0.05). So if we consider Piet Mondrian's 1920-1940 oeuvre as obtained by sampling an unknown ideal design space of worthwhile black-lines-colored-planes compositions, then the median of the splittingness (in this design space) cannot be 1.0 whereas 0.5 is a serious possibility. Plot 1 shows how the SPLITTINGNESS (*percentage of Tees not part of a Strange Coincidence*) varies over time. There are some trends, such as a peak around 1922 (lots of splitting), and a stabilizing phase around 20% after 1936 (many crossings).

RQ2: for the *simple complexity measure* COMPLEXITY we find values between 1.13 and 6.63. Plot 2 shows how the *simple complexity measure* varies over time. In the period 1920-1925 it goes down from about 5 to about 2. After B160 (1926) a stable low-complexity period begins, which gets disrupted around B258 (1935) and then increases graduallyRQ3: for the *special effects* we find numbers in the range from 0 (B104, many more) and 9 (B130). After 1927 (B189) this number is zero. Of all paintings, 35 out of 150 are non-zero i.e. 23%.

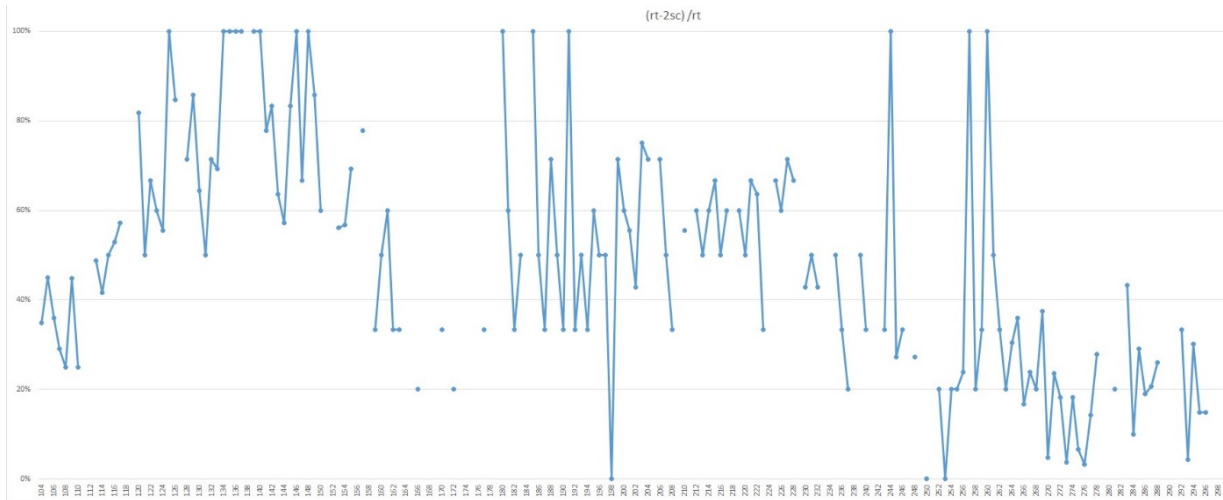

**Plot 1:** Percentage of Tees supportive of the hypothesis of splitting decomposition

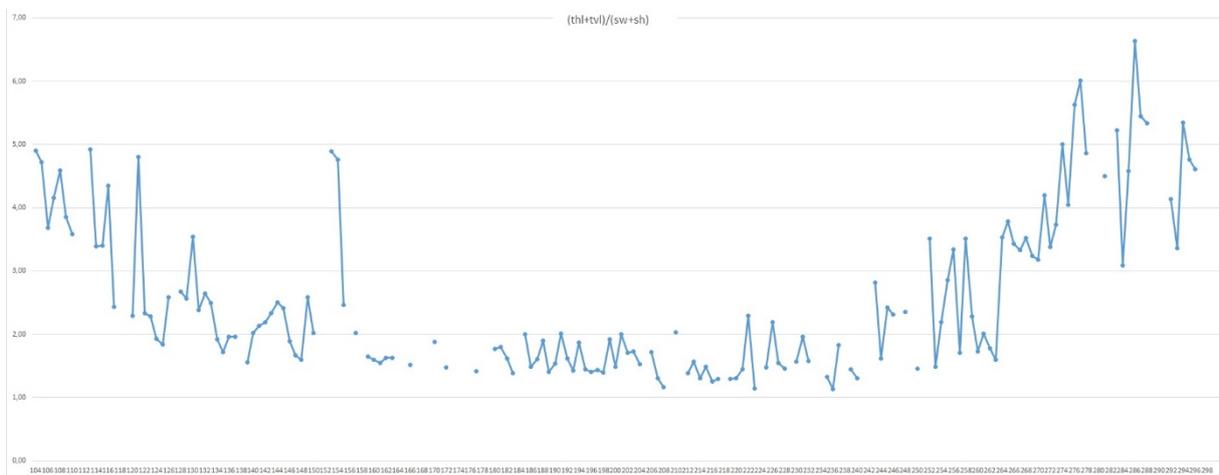

**Plot 2:** Simple complexity measure plotted for the included paintings from 1920 till 1940.

## Summary, Discussion and Options for future work

Our main conclusion is that the hypothesis of splitting decomposition is in general not true. It is possible to make Mondrian-like compositions by splitting, yet one misses out on a great deal of Neoplastic beauty if one would work by splitting only. It is possible to consider all crossings as pairs of Tees, but that is clumsy, and it leaves out essential information. Moreover there are other important design decisions of Mondrian, such as the keeping-distance to the canvas-edge, which are not well-described by splitting.

The application of the Wilcoxon signed rank test deserves a discussion. Whether the mentioned unknown *ideal design space* exists is speculative. The concept is neo-Platonic, but so is the Neoplastic standpoint that the artist has the task of finding expression for pure aesthetic ideas[5]. The Wilcoxon signed rank test does not require the data to be normally distributed, but it *does* require the samples to be taken independently and randomly. We do not imagine reducing Mondrian to a random sampler, he knew what he was doing. But for the statistics, his process perhaps worked like Wolfram's rule 30 in the sense that rule 30 is deterministic, yet for statistical purposes behaves as if it were random. Possibly the sampling was not independent (indeed, we noted trends), but we argue that this is no problem since we re-tested the hypothesis that the median splittingness is 100% in the high-splitting subrange B125-B148 (20 paintings, see Plot 1) and found by the same One-Sample Wilcoxon Signed Rank Test that even this hypothesis must be rejected (significance level is 0.05).

---

[5] In Mondrian's own words: "Z: [] we ought to see more deeply, our vision should be abstract, universal. Then externality will become for us what it really is: the mirror of truth." In (Mondrian 1919) p.310

The analysis itself has its weak spots, notably the arbitrariness of the decisions taken during the marking process. To our defense, we like to remark that if the splitting hypothesis was fully true (all paintings being 100% splitting-based), there would not be any arbitrariness. There is evidence that the ambiguity of how to "see" the work was an essential part of Mondrian's design repertoire. It would be interesting to explore a quaternary splitting hypothesis (the present hypothesis being binary). The simple complexity measure is probably too simple, an entropy-based method could be promising (Marsden 2020). There are more weaknesses of the present analysis, as we did not consider the colored planes, the colors, the role of double lines, the super-planes, not even the widths of the lines. As another option for future work, it would be interesting to translate the gained insights into a new Mondrian-like composition generator.

Finally, as a last paragraph of tribute to Mondrian and as a confirmation of our findings we show a photograph of the 1965 Mondrian-like dresses designed by Yves Saint Laurent (1936-2008). This is one of the many examples of the impact of Mondrian in the world. The 1965 fashion collection of Yves Saint Laurent is an iconic collection by one the most admired fashion designers of all times. In Figure 3 we see three dresses, two of which having crossings only (also on the back, although we cannot see that in the photo), whereas one is of the mixed splitting-crossing type. It is plausible that Yves Saint Laurent has drawn a conclusion similar to ours. Note the painting in the background, B116, Composition with yellow, blue, black, red, and gray (1921), which according to our analysis has a *splittingness* of 53%, a *complexity* of 4.35, and a *special effect* count of 3. In the 2014 biographic movie on Yves Saint Laurent directed by Jalil Lespert, there is a shot of Yves looking for inspiration; going through a Mondrian book his eye is caught by B108, *Composition I* (1920), having 25%, 4.59 and 2, respectively. Then he begins sketching for the iconic collection. Later, Yves and Pierre Berger bought B142 *Composition avec bleu, rouge, jaune et noir*, one pre-1920 work, and B108 (B142 was sold by Berger in 2009 for €21,569,000).

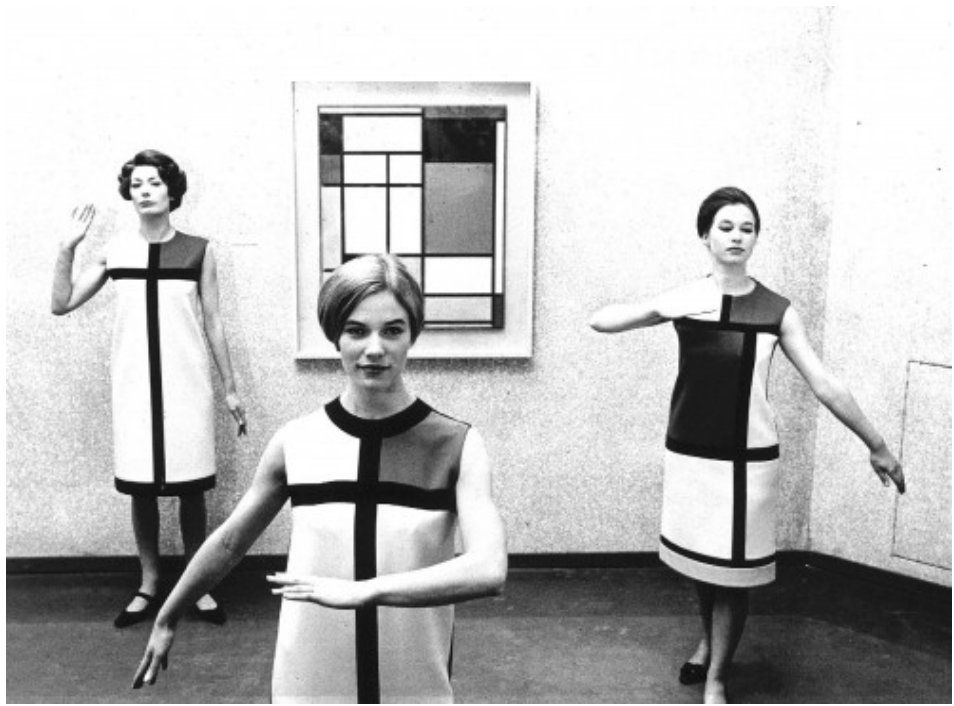

**Figure 3:** The 1965 designs of Yves Saint-Laurent. www.kunstmuseum.nl/en/exhibitions/fashion-style (Three colleagues of the former Dutch Costume Museum dressed in Yves Saint Laurent dresses, fall 1965, in the Kunstmuseum Den Haag, 1966)

Acknowledgements: we are grateful for the work by Joosten and Welsh and RKD, the Netherlands Institute for Art History, for the development and maintenance of the Catalogue Raisonné and the pietmondrian.rkdmonographs.nl website, which is a great resource for Mondrian research.